\documentclass[a4paper,11pt]{article}
\usepackage{rotating}
\usepackage{jheppub} 
\usepackage[utf8]{inputenc}
\usepackage{placeins} 
\usepackage{chngpage}
\usepackage{amsmath,amsfonts}
\usepackage{graphicx}
\usepackage{color}
\usepackage{xcolor}
\usepackage{relsize}
\usepackage{epstopdf}
\usepackage{hyperref}
\usepackage{mathrsfs}
\usepackage{ragged2e}
\usepackage{amssymb}
\usepackage{placeins}
\usepackage[normalem]{ ulem }
\usepackage{amsthm}
\usepackage{comment}
\usepackage{graphics}
\usepackage{caption}
\usepackage{subcaption}
\usepackage{verbatim}
\usepackage{arydshln}
\usepackage{cprotect}
\usepackage{adjustbox}

\usepackage{bbold}

\usepackage{tabularx,booktabs}
\usepackage{multicol}
\usepackage{array,multirow}
\usepackage{booktabs}
\usepackage{colortbl}
\usepackage{float}
\def\gsim{\raise0.3ex\hbox{$\;>$\kern-0.75em\raise-1.1ex\hbox{$\sim\;$}}}
\def\lsim{\raise0.3ex\hbox{$\;<$\kern-0.75em\raise-1.1ex\hbox{$\sim\;$}}}

\makeatletter
\gdef\@fpheader{}
\vspace*{0cm}
\makeatother

\begin{document}

\title{Faking ZZZ vertices at the LHC}

\author[a]{Ricardo Cepedello,}
\author[b]{Fabian Esser,}
\author[b]{Martin Hirsch}
\author[b]{and Veronica Sanz}
\affiliation[a]{Departamento de Fisica Teorica y del Cosmos, 
Universidad de Granada, Campus de Fuentenueva, E-18071 Granada, Spain}
\affiliation[b]{Instituto de F\'isica Corpuscular (IFIC), 
Universidad de Valencia-CSIC, E-46980 Valencia, Spain}

\emailAdd{ricepe@ugr.es}
\emailAdd{esser@ific.uv.es}
\emailAdd{mahirsch@ific.uv.es}
\emailAdd{veronica.sanz@uv.es}

\date{\today}% It is always \today, today,
             % but any date may be explicitly specified 

\abstract{ 

Searches for anomalous neutral triple gauge boson couplings (NTGCs)
provide important tests for the gauge structure of the standard
model. At the LHC, NTGCs are searched for via the process $pp \to ZZ
\to 4l$, where the two $Z$-bosons are on-shell. In this paper, we
discuss how the same process can occur through tree-level diagrams
just adding a vector-like quark (VLQ) to the standard model. Since
NTGCs are generated in standard model effective theory (SMEFT) only at
1-loop order, vector like quarks could be an important alternative
interpretation to, and background for, NTGC searches.  Here, we
construct a simple example model, discuss low-energy constraints and
estimate current and future sensitivities on the model parameters from
$pp \to ZZ \to 4l$ searches.

}

\keywords{SMEFT, UV completions, LHC physics, precision observables}
             
\maketitle

\section{Introduction\label{sec:intro}}

Neutral triple gauge boson couplings (NTGCs) provide an important test
for the gauge structure of the standard model (SM). Both ATLAS
\cite{ATLAS:2017bcd} and CMS \cite{CMS:2020gtj} have searched for
triple-$Z$ couplings ($ZZZ$) using the four lepton final state. This
final state is more sensitive to new physics than hadronic decay modes
of the $Z$, despite the small branching ratio of $Z$-bosons to charged
leptons. It is important to note that NTGC searches are currently not
background limited, thus considerable improvement in sensitivity can
be expected for the high-luminosity phase of the LHC.

From a theoretical point of view, however, generating NTGCs of
observable size seems difficult. In SMEFT (``Standard Model Effective
Field Theory'') NTGCs appear first at the level of $d=8$ operators
\cite{Gounaris:2000dn,Degrande:2013kka}, thus cross sections for NTGCs
at fixed $\sqrt{s}$ scale as $\sigma \propto (v/\Lambda)^8$. There are
in total four CP-conserving $d=8$ operators that contribute to NTGCs~\footnote{Recently, ref.~\cite{Ellis:2024omd}
has shown that three additional operators contribute to off-shell
production. However, since off-shell production is highly suppressed,
the sensitivity to these additional operators is also limited.} \cite{Cepedello:2024ogz} and
all of these $d=8$ operators can only be generated at 1-loop level
\cite{Gounaris:2000tb,Cepedello:2024ogz}, leading to an additional
$(16\pi^2)$ suppression in the vertices.\footnote{There are also CP
violating operators contributing to NTGCs. However, here theory
expectations are even more pessimistic. For example, for the
Two-Higgs-Doublet Model it has been shown in
\cite{Belusca-Maito:2017iob} that CPV NTGCs appear dominantly at
1-loop and $d=12$.}

\begin{figure}[t!]
    \centering
    \includegraphics[scale=0.85]{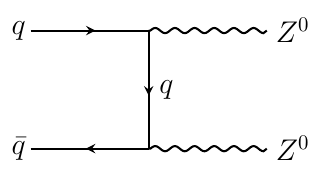}\hskip1mm
    \includegraphics[scale=0.85]{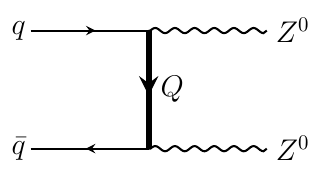}\hskip1mm
    \includegraphics[scale=0.85]{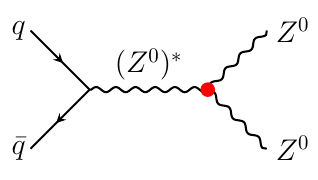}
    \caption{Standard model (left) and vector-like quark (middle)
      contributions to ${\sigma(pp \to ZZ)}$ at the LHC. To the right,
      NTGC contribution. For large values of the vector-like quark
      mass the diagram scales like a $d=8$ operator, exactly like
      the NTGC vertex diagram, see text.}
    \label{fig:diags}
\end{figure}

The main idea of the current paper is that the same final state as
used in the NTGC search can be easily generated just adding a heavy
vector-like quark to the standard model.  Fig.\ \ref{fig:diags} shows
two tree-level diagrams that generate two on-shell $Z$-bosons at the
LHC. The diagram on the left is the standard model background for the
$ZZZ$ search, while the diagram in the middle adds a vector-like
quark, mixed with SM quarks, to the SM particle content.  In addition,
a NTGC diagram with a triple-Z vertex is shown on the right in
Fig.\ \ref{fig:diags}. Here, $q$ can stand for either $q=u,d$.
Since, as discussed above, the vertex in the NTGC diagram can be
generated only at 1-loop order, the tree-level VLQ diagram can
potentially give much larger rates and can thus be an important
background to NTGC searches. 

Since on the energy scale of the LHC the SM quarks can be considered
massless, the invariant mass $m_{ZZ}$ of the $ZZ$ system from the SM
diagram, Fig.\ \ref{fig:diags}, peaks at roughly $m_{ZZ}=2 m_Z$ and
the differential distribution drops rapidly at larger $m_{ZZ}$. The
NTGC diagram, on the other hand, corresponds to a $d=8$ SMEFT operator
and thus peaks at larger values of $m_{ZZ}$.  Experimentalists
therefore cut on $m_{ZZ}$ to reduce backgrounds when searching for a
NTGC vertex.

The diagram in the middle in Fig.\ \ref{fig:diags}, however, is
different from the SM diagram. The flavour violating vertices are
mixing suppressed, i.e.\ each vertex in the diagram is proportional to
$(y v)/m_Q$, where $y$, $v$ and $m_Q$ are some Yukawa coupling, the
standard model vacuum expectation value (vev) and the mass of the
heavy VL quark, respectively. The propagator of the heavy quark is
proportional to $1/m_Q^2$. Thus, in the limit of large VL quark mass,
the diagram will in total be proportional to $1/m_Q^4$ -- exactly the
same as a $d=8$ operator.

\begin{figure}[t!]
    \centering
    \includegraphics[scale=1.0]{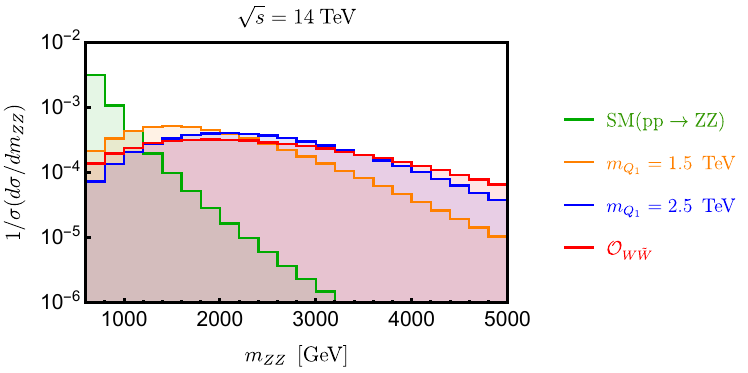}
    \caption{Differential cross section $\sigma(pp \to ZZ)$ as
      function of the invariant mass $m_{ZZ}$ at the 14 TeV
      LHC. Shown are the standard model contribution, the
      distributions for the VLQ diagram in Fig.\ \ref{fig:diags} for
      two mass choices, and the case of one particular NTGC operator,
      ${\cal O}_{W{\tilde W}}$.}
    \label{fig:SigmZZ}
\end{figure}

Fig.\ \ref{fig:SigmZZ} shows the differential cross section
$\sigma(pp \to ZZ)$ as function of the invariant mass $m_{ZZ}$ for
four different example calculations. Shown are the SM distribution,
the expected distribution for one specific NTGC operator ${\cal
  O}_{W{\tilde W}}$ (see App.\ \ref{sec:appNTGC}), and two
distributions for a model with a vector-like quark corresponding to
two mass options of 2.5 and 1.5 TeV. The distribution for $m_{ZZ}$ for
the EFT operator and the heavy mass VLQ calculations are very similar
and will thus be difficult to distinguish experimentally. We call this
``faking a NTGC''. Clearly, since the VLQ model generates this cross
section at tree-level, while ${\cal O}_{W{\tilde W}}$ can only be
generated at 1-loop, the naive expectation is that the VLQ model can
give an important background to the NTGC searches.  The rest of this
paper is dedicated to work out this claim quantitatively.

This paper is organized as follows: In section \ref{sec:vlqs} we
briefly discuss vector-like quarks. We give some basic definitions for
vector-like quarks in \ref{sec:mod} and examine constraints on the
model parameters from a variety of searches, both LHC and low-energy,
in section \ref{sec:cnst}. Section \ref{sec:pheno} then discusses the
cross section calculation and gives estimates for current and future
LHC sensitivities. We close in section \ref{sec:cncl} with a short
discussion and conclusion.

\section{Vector-like quarks\label{sec:vlqs}}

In this section we introduce models with heavy vector-like quarks that
can generate final states with two on-shell Z bosons. We discuss their
mass mixing with the SM quarks and their contribution to charged and
neutral currents. Finally, we use low-energy observables to constrain
the masses and Yukawa couplings of these models.

\subsection{VLQ Lagrangian and interactions\label{sec:mod}}
In total there are seven different vector-like quarks (VLQs) with
couplings to a SM quark and the Higgs. These are: Two singlets
($U_R=F_{3,1,2/3}$, $D_R=F_{3,1,-1/3}$), three doublets
($Q_1=F_{3,2,1/6}$, $Q_5=F_{3,2,-5/6}$ and $Q_7=F_{3,2,7/6}$) and two
triplets ($T_1=F_{3,3,-1/3}$ and $T_2=F_{3,3,2/3}$). \footnote{Indices
denote transformation properties or quantum numbers under the SM gauge
group, $SU(3)_C\times SU(2)_L\times U(1)_Y$.}  The Lagrangian
involving the heavy fields can be written as:
\begin{equation}\label{eq:lag}
{\cal L}  = {\cal L}_{\rm kin} +  {\cal L}_{\rm Y} +  {\cal L}_{\rm M}
             + {\cal L}_{\rm H}
\end{equation}
${\cal L}_{\rm kin}$ are the kinetic terms, while ${\cal L}_{\rm Y}$ 
contains the Yukawa interactions:
\begin{eqnarray}\label{eq:lagY}
{\cal L}_{\rm Y} &=& Y_{U_R} \overline{Q} U_R H^{\dagger}
          +  Y_{D_R} \, \overline{Q} D_R H
          +  Y_{Q_u} \, \overline{Q_1} u_R H^{\dagger}
          +  Y_{Q_d} \, \overline{Q_1} d_R H \\ \nonumber
         & + & Y_{Q_7} \, \overline{Q_7} u_R H
          +  Y_{Q_5} \, \overline{Q_5} d_R H^{\dagger}
          +  Y_{T_2} \, \overline{Q} T_2 H^{\dagger}
          +  Y_{T_1} \, \overline{Q} T_1 H + {\rm h.c.}
\end{eqnarray}
The VLQ mass terms are:
\begin{eqnarray}\label{eq:lagM}
{\cal L}_{\rm M} &=& m_{U_R} \, \overline{U_L} U_R 
                +  m_{D_R} \, \overline{D_L} D_R 
                +  m_{Q_1} \, \overline{Q_{1,L}} Q_{1,R} 
                +  m_{Q_5} \, \overline{Q_{5,L}} Q_{5,R} \\ \nonumber
             &  + &  m_{Q_7} \, \overline{Q_{7,L}} Q_{7,R} 
                +   m_{T_1} \, \overline{T_{1,L}} T_{1,R} 
                +  m_{T_2} \, \overline{T_{2,L}} T_{2,R} 
                + {\rm h.c.}
\end{eqnarray}
It is important to note that for singlet and triplet VLQ states the
Yukawa interactions connect them to the SM doublet $Q$.  For the VLQ
doublets, however, the Yukawa couplings connect these states to the SM
singlets $u_R$ and $d_R$. All differences in phenomenology between
singlet and triplets on one hand and doublet VLQs on the other hand
can be traced back to this distinct nature of the couplings.
${\cal L}_{\rm H}$ stands for a possible set of Yukawa couplings
connecting two VLQs to the SM Higgs. In the following, we will neglect
these terms for simplicity.

For reasons explained in detail below, we will concentrate on one
particular subset of VLQs, namely $Q_1 = F_{3,2,1/6}$ and $Q_7 =
F_{3,2,7/6}$, in our numerical analysis.  We will comment on
similarities and differences with other VLQs in passing in the text.
With the new terms in eq.\ (\ref{eq:lag}) the mass matrices for up and
down quarks are changed relative to the SM definitions.  In the
following, SM quarks will be assigned lower case letters, while we
will denote the heavy states as $D_1$, $U_1$ and $U_7$, where the
formers are the two components of the doublet $Q_1$, while the latter
is the lowest isospin $SU(2)_L$ component of $Q_7$. In the basis
$F_{u_L}=\{u_L,U_{1,L},U_{7,L}\}$ versus ${\bar F_{u_R}}=\{{\bar u_R},
{\bar U_{1,R}},{\bar U_{7,R}}\}$ and $F_{d_L}=\{d_L,D_{1,L}\}$ versus
${\bar F_{d_R}}=\{{\bar d_R}, {\bar D_{1,R}}\}$, the full quark mass
matrices can be written as:
\begin{eqnarray}\label{eq:MassUp}
  {\cal M}_{d} =
  \begin{pmatrix}
    \frac{1}{\sqrt{2}}y_dv & 0 \\ 
    \frac{1}{\sqrt{2}}Y_{Q_d}v & m_{Q_1} \\
  \end{pmatrix},
& \hskip10mm &  {\cal M}_{u} =
  \begin{pmatrix}
    \frac{1}{\sqrt{2}}y_uv & 0 & 0\\ 
    \frac{1}{\sqrt{2}}Y_{Q_u}v & m_{Q_1} & 0 \\
    -\frac{1}{\sqrt{2}}Y_{Q_7}v & 0 & m_{Q_7} \\
  \end{pmatrix}.
\end{eqnarray}  
The SM part ($y_u$, $y_d$) is to be understood as a $3 \times 3$
matrix, whereas for the heavy states we assume for simplicity only one
generation of $Q_1$ and $Q_7$. Thus, $Y_{Q_\alpha}$ for $\alpha=u,d,7$
are vectors with three components. The mass matrices can be
bi-diagonalized via:
\begin{equation}\label{eq:bidi}
\hat{\cal M}_u = V_u^{\dagger} {\cal M}_u U_u,
\end{equation}  
and similarly for ${\cal M}_d$. Non-zero values of the Yukawa
couplings $Y_{Q_u}$, $Y_{Q_7}$ and/or $Y_{Q_d}$ will affect both the
mixing matrices and the light eigenvalues.  Light eigenvalues are
shifted with respect to their SM values, which can be easily corrected
for by re-adjusting the SM Yukawa couplings, at leading order:
\begin{equation}\label{eq:shiftY}
  y_{x_i} \simeq y_{x_i}^{\rm SM} \times
            \left[ 1 - \left(\frac{Y_{Q_\alpha}^{(i)}v}{2 m_{Q_\alpha}}\right)^2 \right]^{-1},
\end{equation}  
where $i$ stands for the $i$th SM generation and $x=u,d$. The left-
and right rotation matrices, $V_{u/d}$ and $U_{u/d}$,
take a particularly simple form if only one of the components of
$Y_{Q_x}^{(i)}$ is non-zero. The angles in the left- and right-mixing
matrices are then given in leading order as
\begin{eqnarray}\label{eq:MixApp}
s_{i4,d}^{V}  \simeq  \frac{Y_{Q_d}^{(i)}v }{\sqrt{2}m_{Q_1}} 
\Big(\frac{m_{d_i}}{m_{Q_1}}\Big),
& \hskip10mm &
s_{i4,u}^{V}  \simeq  \frac{Y_{Q_{u/7}}^{(i)}v }{\sqrt{2}m_{Q_{1/7}}} 
\Big(\frac{m_{u_i}}{m_{Q_{1/7}}}\Big),
\\ \label{eq:MixAppR}
s_{i4,d}^{U}  \simeq  \frac{Y_{Q_d}^{(i)}v }{\sqrt{2}m_{Q_1}},
& \hskip10mm &
s_{i4,u}^{U}  \simeq  \frac{Y_{Q_{u/7}}^{(i)}v }{\sqrt{2}m_{Q_{1/7}}},
\end{eqnarray}
where $s_{i4,x}^{V/U}$ is a short-hand for $\sin(\theta_{i4,x}^{V/U})$.
{\bf Left mixing}, proportional to $s_{i4,x}^{V}$, is suppressed by
light quark masses, as eq.\ (\ref{eq:MixApp}) shows, while {\bf right
  mixing} is proportional to $s_{i4,x}^{U}$ and follows the naive
expectation of being $\propto 1/m_Q$.

As it can be seen from the Lagrangian, eq.\ (\ref{eq:lag}), models with
VLQ singlets or triplets would couple to SM doublets instead of
singlets, leading to a transposed form of the mass matrices in
eq.\ (\ref{eq:MassUp}). Therefore the expressions for left and right
mixing angles are exchanged with respect to eqs (\ref{eq:MixApp}) and
(\ref{eq:MixAppR}) and the left-mixing is not suppressed by the light
quark masses. As it will be shown in the following, this difference in
the coupling explains why VLQ doublets are much less constrained from
low-energy searches than singlets or triplets.\\ For the following
discussion, it will be useful to give the interactions of quarks with
$W^{\pm}$ and $Z$ bosons. The charged current coupling of quarks in
the model is given by:
\begin{eqnarray}\label{eq:CC}
  {\bar u_i}-d_j-W^+_{\mu} & = &
  -i \frac{g_2}{\sqrt{2}} \left[ \left(
  \sum_{a=1}^{4}V^{d,*}_{j a} V_{i a}^{u}\right) \gamma_{\mu}P_L +
  (U^{u,*}_{i 4} U_{j 4}^{d}) \gamma_{\mu}P_R \right].
\end{eqnarray} 
The up (down) quark index $i$ ($j$) runs from $1$ to $5$ ($1$ to $4$). 
The neutral current couplings can be written in a compact form:
\begin{eqnarray}\label{eq:NC}
  {\bar u_i}-u_j-Z^0_{\mu} = -i \frac{g_2}{c_W}& & \left\lbrace G_L^u
  \left(\sum_{a=1}^{5}V^{u,*}_{j a} V_{i a}^{u}\right) \gamma_{\mu}P_L \right. 
  \\ \nonumber & &  \left. + \left[ G_R^u \left(\sum_{a=1}^{3}U^{u,*}_{j a}U_{i a}^{u}\right)
  -G_L^u U^{u,*}_{j 4} U_{i 4}^{u}+G_L^u U^{u,*}_{j 5} U_{i 5}^{u}\right]
  \gamma_{\mu}P_R \right\rbrace,
\end{eqnarray}
where
\begin{eqnarray}\label{eq:GLR}
  G_L^u = (T_3^u - q^u s_W^2), \\ \nonumber
  G_R^u = - q^u s_W^2 .
\end{eqnarray}
Note that $G_L^u$ for the state corresponding to $U_7$ (index 5, see
eq. (\ref{eq:bidi})) is different from $G_L^u$ of the other states,
since $T_3^{U_7}=-1/2$.  The ${\bar d_i}-d_j-Z^0_{\mu}$ couplings have
very similar expressions, only replacing $G_{L/R}^u$ with the
corresponding $G_{L/R}^d$ and restricting the indices ($i$, $j$) from
$1$ to $4$.  Since left-mixing angles of SM quarks with the heavy
states $U_1$, $D_1$ and $U_7$ are suppressed by light quark masses,
the SM parts of the CC and NC left couplings remain practically
unchanged.  However, there are important changes in the right
couplings.

For the calculation of the constraints on VLQ parameters from low-energy
observables, such as $\beta$- or kaon decays, it is useful to
go one step further and calculate {\em modified} CC and NC couplings
of the SM quarks, once heavy VLQs are integrated out. These CC
and NC vertices can be written as \cite{Kitahara:2024azt}:
\begin{eqnarray}\label{eq:CCNCeff}
  {\cal L}_{W,Z} = -\frac{g_2}{\sqrt{2}}W^+_{\mu} {\bar u_i} \gamma^{\mu}
 & & \Big( [ V_{CKM} ( \mathbb{1} + v^2 C_{Hq}^{3}) ]_{ij} P_L
   + \frac{v^2}{2} [C_{Hud}]_{ij} P_R \Big) d_j + {\rm h.c.} \\ \nonumber
 -\frac{g_2}{6 c_W} Z_{\mu} {\bar u_i} \gamma^{\mu}
  & & \Big( [ (3 - 4 s_W^2)\mathbb{1}
     + 3v^2  V_{CKM} \{ C_{Hq}^{(3)} - C_{Hq}^{(1)}\} V_{CKM}^{\dagger} ]_{ij} P_L
   \\ \nonumber
 & &   - [4 s_W^2 \mathbb{1} + 3 v^2 C_{Hu} ]_{ij} P_R \Big) u_j  \\ \nonumber
 -\frac{g_2}{6 c_W} Z_{\mu} {\bar d_i} \gamma^{\mu}
  & & \Big( [ (2 s_W^2 - 3)\mathbb{1}
     + 3v^2 \{ C_{Hq}^{(3)} + C_{Hq}^{(1)}\} ]_{ij} P_L
   \\ \nonumber
 & &   + [2 s_W^2 \mathbb{1} + 3 v^2 C_{Hd} ]_{ij} P_R \Big) d_j .
\end{eqnarray}
Here, $V_{CKM}$ is the $3\times3$ SM mixing matrix. If we consider only the
contributions from VLQ states $Q_1$ and $Q_7$, the matching of the
Wilson coefficients to the UV parameters is given by:
\begin{eqnarray}\label{eq:mtch}
  [C_{Hu} ]_{ij} &=&- \frac{(Y_{Q_u}^{(i)})^*Y_{Q_u}^{(j)}}{2 m_{Q_1}^2}
                 + \frac{(Y_{Q_7}^{(i)})^*Y_{Q_7}^{(j)}}{2 m_{Q_7}^2},
  \\ \nonumber
  [C_{Hd} ]_{ij} &=& \frac{(Y_{Q_d}^{(i)})^*Y_{Q_d}^{(j)}}{2 m_{Q_1}^2}, \\ \nonumber
  [C_{Hud} ]_{ij} &=& \frac{(Y_{Q_u}^{(i)})^*Y_{Q_d}^{(j)}}{2 m_{Q_1}^2},\\ \nonumber
  [C_{Hq}^{(3)}]_{ij} & = & [C_{Hq}^{(1)}]_{ij} = 0.
\end{eqnarray}  
As discussed in the next subsection, with these equations it becomes
straight-forward to understand why doublet VLQs are much less
constrained from low-energy searches than singlet or triplet
VLQs. Note that for singlet and triplet VLQs both $C_{Hq}^{(3)}$ and
$C_{Hq}^{(1)}$ are non-zero.

\subsection{Constraints on model parameters\label{sec:cnst}}

To study low-energy constraints and calculate cross sections, we have
implemented the model defined by the Lagrangian in eq.\ (\ref{eq:lag})
into \texttt{SARAH}/\texttt{SPheno} \cite{Staub:2013tta,Porod:2011nf},
\texttt{Matchete} \cite{Fuentes-Martin:2022jrf} and also into
\texttt{FeynRules} \cite{Alloul:2013bka}.  The \texttt{SARAH}
implementation was done to cross check the calculation of the Feynman
rules of the model, while the \texttt{FeynRules} implementation allows
us to generate \texttt{UFO} files \cite{Degrande:2011ua}, which are
the input for the calculation of cross sections using
\texttt{MadGraph} \cite{Alwall:2011uj,Alwall:2014hca}. The
\texttt{Matchete} implementation has been used to match the model in
the EFT limit at the level of $d=6$ {\bf and} $d=8$ operators in
SMEFT.

We use the set of $d=6$ operators from the \texttt{Matchete} output,
matched at 1-loop order, to calculate constraints on the model
parameters from a number of observables, both low-energy and LHC.
These constraints are calculated using the \texttt{flavio}
package\footnote{A documentation for \texttt{flavio} and a list of the
included observables can be found on the \texttt{github} page
\textit{https://flav-io.github.io/}}
\cite{Straub:2018kue}. \texttt{Flavio} automatically matches SMEFT
operators to LEFT operators for the calculation of low-energy
variables and includes RGE running. We have checked a number of
observables and will discuss each of them in turn.

First of all, according to the PDG \cite{ParticleDataGroup:2020ssz}
there is a slight anomaly in the determination of the unitarity of the
first row of the CKM matrix, $\Delta_{CKM} = \sum V_{ux}^2 -1 = 0.9985
\pm 0.0007$. A non-zero $\Delta_{CKM}$ can be due to either a change
in $V_{ud}$ and/or $V_{us}$. $\beta$-decays determine the former while
kaon decays are used to determine the latter. From an EFT point of
view, a non-zero $\Delta_{CKM}$ corresponds to a change in the CC
vertices, see eq.\ (\ref{eq:CCNCeff}), and thus is sensitive to
$C_{Hq}^{(3)}$ and $C_{Hud}$
\cite{Falkowski:2017pss,Falkowski:2023klj}. That vector-like quarks
can explain a non-zero $\Delta_{CKM}$ has been discussed in several
papers recently, see for example
\cite{Crivellin:2022rhw,Kirk:2023oez,Cirigliano:2023nol,Kitahara:2024azt}.
Choosing to consider only the VLQ doublets $Q_1$ and $Q_7$, one finds
$\left(C_{Hq}^{(3)}\right)_{ij}=0$, see
Eq.\ (\ref{eq:mtch}). $[C_{Hud}]_{11}$ on the other hand is non-zero
only if both $Y_{Q_d}^{(1)}$ and $Y_{Q_u}^{(1)}$ are non-zero and
$[C_{Hud}]_{12}$ even requires that additionally $Y_{Q_d}^{(2)}$ and
$Y_{Q_u}^{(2)}$ are finite. \cite{Crivellin:2022rhw} claims that such
a choice can fit the $\Delta_{CKM}$ anomaly, consistent with other
constraints. However, we will not attempt to explain $\Delta_{CKM}$.
Disregarding $\Delta_{CKM}$, we can simply set either $Y_{Q_d}$ or
$Y_{Q_u}$ to zero and either scenario is safe from unitarity
constraints, even if the non-zero Yukawa takes arbitrarily large
numerical values.

Next, we mention that constraints from 3rd generation mesons are
numerically not as strong as those from $\beta$ or kaon decays, but
choosing either $Y_{Q_d}$ or $Y_{Q_u}$ zero and following the same
reasoning as before one avoids $B$-meson constraints alltogether.
Also, note that $Q_5$ and $Q_7$ both have only one Yukawa coupling
with the SM quarks. Thus, in a model with these states either
$V_{i4}^u$ ($Q_5$) or $V_{i4}^d$ ($Q_7$) are {\bf automatically} zero
and the right-handed charged current, i.e.\ $C_{Hud}$, vanishes for a
model with only these VLQs.

\begin{figure}[t!]
    \centering
    \includegraphics[scale=0.62]{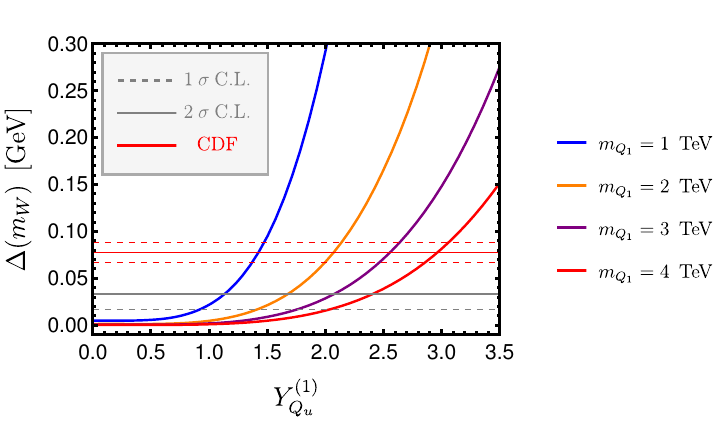}
    \hskip0mm\includegraphics[scale=0.62]{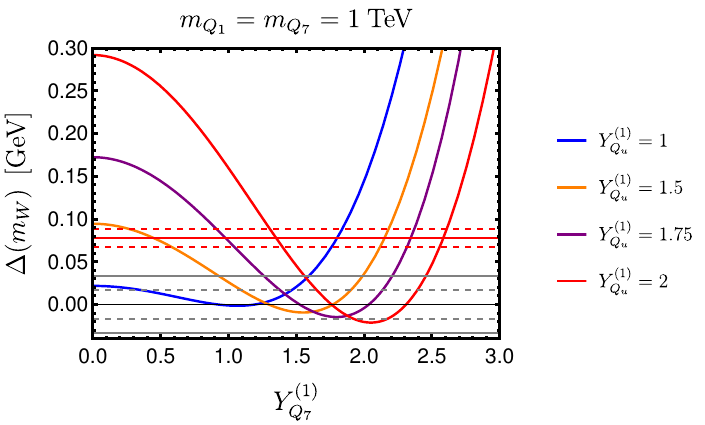}
    \caption{Example calculation for $\Delta(m_W)$ in [GeV] as
      function of $Y_{Q_u}^{(1)}$ for different choices of $m_{Q_1}$
      (left) and as function of $Y_{Q_7}^{(1)}$ for different choices
      of $Y_{Q_u}^{(1)}$ (right). The grey horizontal lines are the
      one and two sigma limits from the global constraints on the
      $W$-boson mass, the red lines are the CDF \cite{CDF:2022hxs}
      measurement (with 1 $\sigma$ error bars).  }
    \label{fig:dmW}
\end{figure}

Furthermore, vector-like quarks, mixed with SM quarks, lead to a shift
in the mass of the W-boson with respect to the standard model
prediction. Fig.\ \ref{fig:dmW} shows an example calculation. Here,
$\Delta(m_W)$ is plotted on the left as function of $Y_{Q_u}^{(1)}$
for several different choices of $m_{Q_1}$ and on the right for
several choices of $Y_{Q_u}^{(1)}$ as function of $Y_{Q_7}^{(1)}$ for a
fixed value of $m_{Q_1}=m_{Q_7}$. In this plot, the grey horizontal
lines are the one and two sigma limits from the global constraints on
the $W$-boson mass \cite{ParticleDataGroup:2020ssz}, while the red
lines are the CDF \cite{CDF:2022hxs} measurement (with 1$\sigma$ error
bars).  It has already been discussed e.g.\ in
\cite{Bagnaschi:2022whn} that a VLQ could explain the CDF
measurement. Such an explanation would require quite large Yukawa
couplings, as the plot shows. On the other hand, even
$Y_{Q_u}^{(1)}=2$ is not excluded for $m_{Q_1}=3$ TeV from the global
SM fit (excluding CDF).
   
Moreover, it is always possible to avoid indirect and low-energy
constraints if there is more than one VLQ and $\Delta(m_W)$ is no
exception to this rule. This is demonstrated by the plot on the right
hand side of Fig.\ \ref{fig:dmW}. Here, large values of
$Y_{Q_u}^{(1)}$ have been chosen and $\Delta(m_W)$ is plotted versus
$Y_{Q_7}^{(1)}$.  For choices of $Y_{Q_u}^{(1)} >1$, $Y_{Q_7}^{(1)}=0$
is not allowed.  However, increasing $Y_{Q_7}^{(1)}$ there is a range
of parameters where the contributions from $Q_7$ and $Q_1$ nearly
cancel each other and in this region there is {\bf no constraint} on
individual Yukawa couplings.

This result can be understood as follows. Several $d=6$ SMEFT
operators contribute to $\Delta(m_W)$, but for a model with VLQs, the
calculation of $\Delta(m_W)$ is dominated by $O_{HD}=(H^{\dagger}
D^{\mu} H)^*(H^{\dagger} D_{\mu} H)$. Considering only contributions
from $Q_1$ and $Q_7$, the matching is given at 1-loop order by
\begin{equation}\label{eq:mtchOHD}
  C_{HD} = - \frac{23 g_1^4}{144 \pi^2\Lambda^2} -
    \frac{g_1^2}{384\pi^2\Lambda^2}(32 |Y_{Q_u}^{(i)}|^2 - 64 |Y_{Q_7}^{(i)}|^2)
    - \frac{64}{512\pi^2\Lambda^2}(|Y_{Q_u}^{(i)}|^2 -  |Y_{Q_7}^{(i)}|^2)^2
\end{equation}  
For Yukawas order ${\cal O}(1)$ or larger, the last term dominates
and $C_{HD}$ nearly cancels when $Y_{Q_u}^{(1)} \sim Y_{Q_7}^{(1)}$.
Because of the contributions from the first two terms in
eq. (\ref{eq:mtchOHD}) an exact cancellation does not occur exactly
at $Y_{Q_u}^{(1)} = Y_{Q_7}^{(1)}$, but at two particular values for
$Y_{Q_7}^{(1)}$ for any fixed $Y_{Q_u}^{(1)}$. Thus, constraints from
$\Delta(m_W)$ can be avoided completely at the price of adding
two new states to the SM.
We mention that there are several other combinations of VLQs that
allow the same cancellation in $\Delta(m_W)$, for example the
combination $Y_{Q_d}$ and $Y_{Q_5}$ or $Y_{D_R}$ and $Y_{T_1}$.

\begin{figure}[t!]
    \centering
    \includegraphics[scale=0.47]{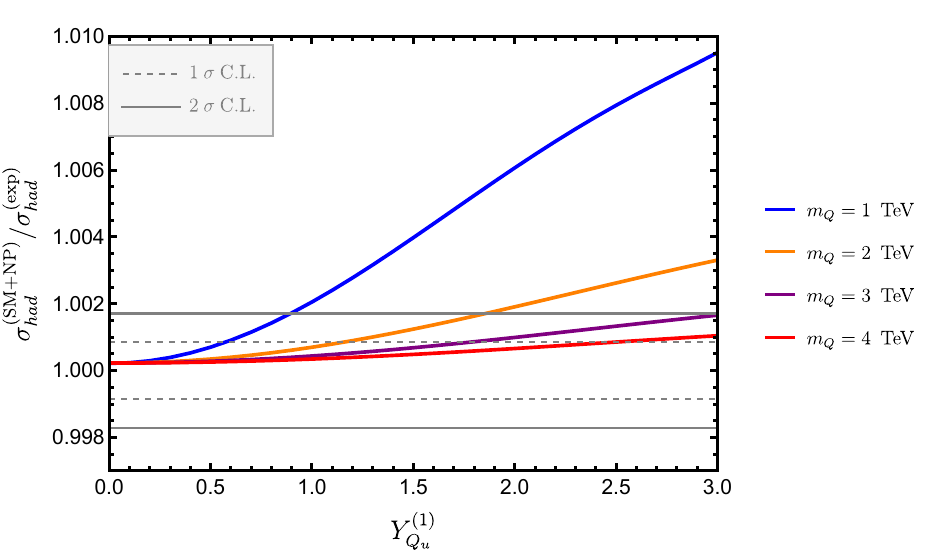}
    \hskip0mm\includegraphics[scale=0.47]{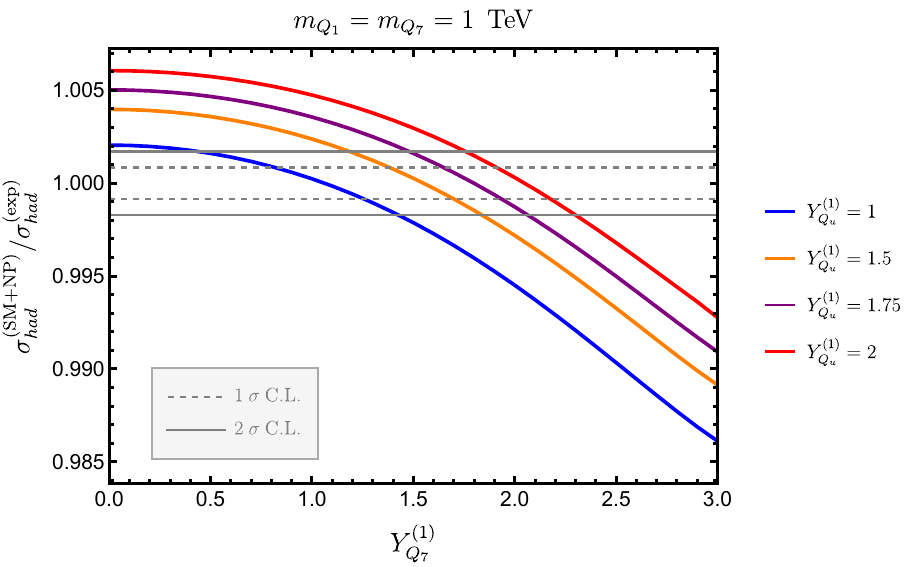}
    \caption{New physics plus SM prediction for $\sigma_{had}$
      normalised to the measured value as a function of
      $Y_{Q_u}^{(1)}$ for different choices of $m_Q$ (left) and as a
      function of $Y_{Q_7}^{(1)}$ for different values of $Y_{Q_u}^{(1)}$
      (right). The grey horizontal lines are the one and two sigma
      limits from the combined experimental and theory prediction
      uncertainty.}
    \label{fig:sighad}
\end{figure}

Finally, LEP has measured very precisely all decay channels of the
$Z^0$ boson. Among the LEP observables is $\sigma_{had}$, the cross
section for $e^+ e^- \rightarrow Z^0 \to$ hadrons at the $Z^0$ pole,
which is particularly precise.  Fig.\ \ref{fig:sighad} shows the SM+NP
prediction normalised to the measured value on the left-hand side for
a model with only $Q_1$ as a function of $Y_{Q_u}^{(1)}$ for different
values of the mass $m_{Q_1}$ and on the right-hand side for a second
model with $Q_1$ and $Q_7$ as a function of $Y_{Q_7}^{(1)}$ for
different values of $Y_{Q_u}^{(1)}$. In the latter case both particle
masses are fixed to 1 TeV. The dashed and solid grey line show the 1
and 2 sigma limits, respectively, where the total uncertainty contains
the experimental uncertainty, the statistical uncertainty in the
calculation of the SM prediction and the deviation of the SM
prediction from the measurement. The first plot shows that for $m_Q=1$
TeV Yukawa couplings $Y_{Q_u}^{(1)}$ larger than ${\cal O}(1)$ are
excluded by the experimental data, but for larger masses these limits
become weaker. As can be seen in the right-hand side, an exact
cancellation between the two new states occurs for $Y_{Q_u}^{(1)} =
Y_{Q_7}^{(1)} =1$. This is the case because the dominant contribution
to $\sigma_{had}$ comes from the SMEFT operator $C_{Hu}$ which is
generated by both $Y_{Q_1}$ and $Y_{Q_7}$ in the matching process,
cf.\ eq.\ (\ref{eq:mtch}). For $Y_{Q_u} = Y_{Q_7}$ and $m_{Q_1} =
m_{Q_7}$ their contributions cancel identically. Furthermore, for a
non-zero $Y_{Q_7}^{(1)}$ larger values of $Y_{Q_u}^{(1)}$ are allowed
due to the partial cancellation. As for the case of $\Delta(m_W)$, no
limits on individual couplings $Y_{Q_u}$ and $Y_{Q_7}$ can be derived
in this cancellation region.
 
In summary, we have discussed a number of constraints on VLQ parameters
from both low-energy and accelerator precision data. A model
with $Q_1$ in which either only $Y_{Q_u}^{(1)}$ or $Y_{Q_d}^{(1)}$ is
non-zero is only weakly constrained at the moment. For a model with
two VLQ states $Q_1$ and $Q_7$ there are even regions in the parameter
space with $Y_{Q_u} \sim Y_{Q_7}$, where the only limit on Yukawa
couplings come from non-perturbativity arguments.

\section{Phenomenology at the LHC\label{sec:pheno}}

In the preceding sections, we covered the theoretical framework for
describing $ZZ$ final states that could mimic non-standard triple
gauge couplings (NTGCs), specifically within models that include new
vector-like quarks. We noted that these models are currently only
weakly constrained by precision measurements, such as the $W$ boson
mass. In this section, we will examine the current and prospective
sensitivity of the Large Hadron Collider (LHC) to these models. We
will begin by studying the potential for direct production of these
new resonances, considering both single and pair production
modes. Following that, we will discuss the indirect constraints that
can be derived from $ZZ$ search results.

\subsection{Pair and single VLQ production at the LHC}

\begin{figure}[t!]
    \centering
    \includegraphics[scale=0.7]{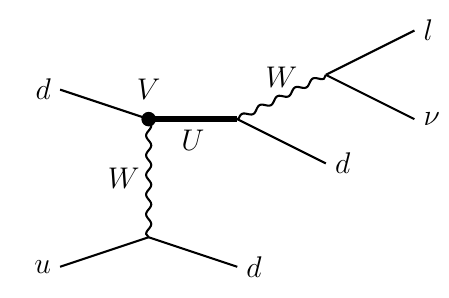}\hskip5mm
    \includegraphics[scale=0.7]{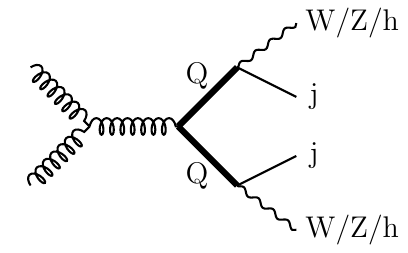}
    \caption{Examples of Feynman diagrams for single and double
      production of VLQs.}
    \label{fig:directprod}
\end{figure}

We begin by discussing the direct searches for VLQs. These new states
can be produced singly, as shown in the left panel of
Fig.~\ref{fig:directprod}, or through strong pair production. The
decays of these particles into an electroweak state ($W/Z/h$) and a
jet would result in a diverse range of final states

The VLQ doublet $Q_1$ can be constrained by direct searches at the
LHC. While most LHC searches concentrate on VLQ decays to 3$^{\rm rd}$
generation SM fermions, ATLAS has recently also published a search for
pair-produced VLQs decaying to 1$^{\rm st}$ generation quarks
\cite{ATLAS:2024zlo}. This search uses the decay $Q \to W + j$ and
requires that one of the two $W$ decays leptonically, in order to
reduce backgrounds. Limits are provided as function of the VLQ mass
and as function of Br($Q \to W + j$) versus Br($Q \to H + j$),
assuming Br($Q \to W + j$)+Br($Q \to H + j$)+Br($Q \to Z + j$)$=1$
\cite{ATLAS:2024zlo}.

\begin{figure}[t!]
    \centering
    \includegraphics[scale=0.7]{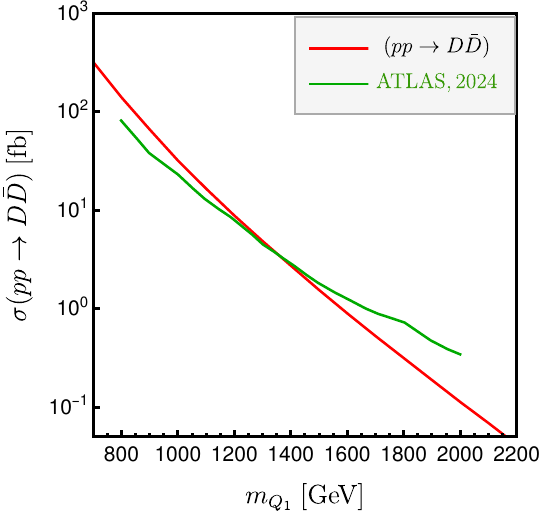}\hskip5mm
    \includegraphics[scale=0.7]{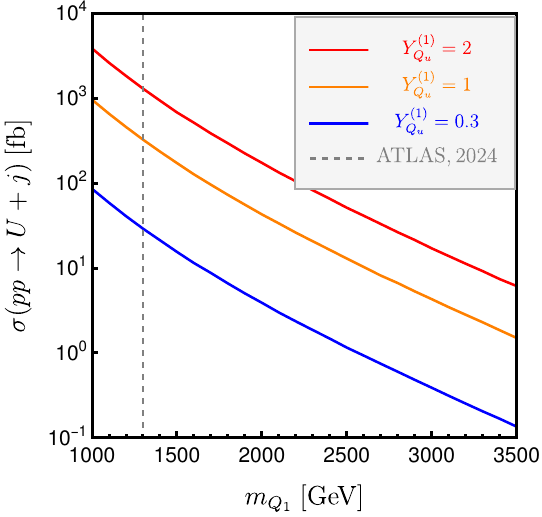}
    \caption{LHC cross sections for (i) pair production $\sigma(pp \to
      D{\bar D})$ at $\sqrt{s}=13$ TeV (left) and (ii) single $U$
      production, $\sigma(pp \to U + j)$ at $\sqrt{s}=14$ TeV (right)
      as function of $m_{Q_1}$.  For pair production ATLAS has
      recently published limits for VLQs decaying to first generation
      quarks and a $W^{\pm}$ \cite{ATLAS:2024zlo}. This limit on the
      production cross section gives roughly a lower limit on
      $m_{Q_1}$ of $m_{Q_1} \gsim 1.3$ TeV.}
    \label{fig:Xsect}
\end{figure}

For $Q_1$ one can show easily, using eqs.\ (\ref{eq:CC}) and
(\ref{eq:NC}), that $U$ will decay to $Z+u$ and $h+u$, if
$Y_{Q_d}\equiv 0$ and $Y_{Q_u}^{(1)}\ne 0$, while it will decay to
$W^+ +d$ if $Y_{Q_u}\equiv 0$ and $Y_{Q_d}^{(1)}\ne 0$. For $D$ the
decay pattern is exactly the opposite. In other words, depending on
the choice which Yukawa coupling is non-zero, either $U$ or $D$ will
be constrained by the search \cite{ATLAS:2024zlo} (or both if both
$Y_{Q_d}$ and $Y_{Q_u}$ are non-zero).  Cross sections for pair $D$
production at the LHC are shown in Fig.\ \ref{fig:Xsect}, left. This
calculation assumes $Y_{Q_d}\equiv 0$, thus $D$ decays to $W+j$ with a
branching ratio close to 100 \%.  As the figure shows, the ATLAS
search puts a lower limit of $m_{Q_1} \gsim 1.3$ TeV. We note that this is
only a rough estimate, since we do not take into account any
$k$-factors in the cross section calculation. Very similar numbers can
be derived for the case $Y_{Q_u}\equiv 0$ with $Y_{Q_d}^{(1)}\ne 0$,
since pair $D$ and pair $U$ production cross sections are very
similar.

In what follows below, we will want to use large values of the Yukawa
coupling $Y_{Q_u}^{(1)}$, in order to maximize $\sigma(pp\to ZZ)$.
Fig.\ \ref{fig:Xsect}, on the right shows the cross sections for
$\sigma(pp\to U +j)$ for different choices of $Y_{Q_u}^{(1)}$. Single
$U$ production cross sections are proportional to $\propto
(Y_{Q_u}^{(1)})^2$ and thus are larger than pair production cross
sections for Yukawa couplings order ${\cal O}(1)$. Moreover, single
$U$ cross sections decrease much more slowly as a function of VLQ mass
than pair production cross sections. Thus, in principle, one might
expect that a search for singly produced $U$ could derive stronger
limits on $m_{Q_1}$, if $Y_{Q_u}^{(1)}$ is sufficiently large.

Both ATLAS and CMS have indeed searched for single VLQ
production. However, in the search \cite{ATLAS:2022ozf} ATLAS assumes
that the VLQ decays to the final state $t+h$ or $W + b$, where the
top/bottom is tagged in order to reduce backgrounds. Similarly CMS
\cite{CMS:2024qdd} searched for single VLQ production in the final
state $t+h$ and $t+Z$. Thus, if only the 1$^{\rm st}$ Yukawa couplings
are non-zero (or at least$Y_{Q_u}^{(3)} \ll Y_{Q_u}^{(2)},Y_{Q_u}^{(1)}$), $U$
decays to tops/bottoms are zero (have very small branching ratios) and
neither the ATLAS \cite{ATLAS:2022ozf} nor the CMS \cite{CMS:2024qdd}
search will provide limits in this case.

A single VLQ search relying on the final states $(W/Z/h)+j$ 
certainly will suffer much larger backgrounds than searches using third
generations, such as \cite{ATLAS:2022ozf,CMS:2024qdd}. However, given the
comparatively large cross section, especially at large $m_Q$, and the
fact that $\sigma(pp \to ZZ) \propto (Y_{Q_u}^{(1)})^4$, a single VLQ
search could extend the mass reach of LHC experiments to
$m_{Q_1} \ge 3.5$ TeV for $Y_{Q_u}^{(1)} \gsim 1$.

In the case of a single VLQ, the signal would be $(W/Z/h)$ plus 2
energetic jets, where one of the jets would be recoiling against the
other jet and the vector boson, daughters of the VLQ decay. Searches
for this final state, massive gauge bosons plus jets, are performed at
LHC experiments, and usually geared towards vector boson fusion (VBF)
production. But the typical VBF kinematical situation, forward jets
with high invariant mass, is not necessarily the best option for VLQ
single production.

For example, in the CMS search \cite{CMS:2019nep} for a leptonic $W$
boson with two jets, a pre-selection on $m_{jj}>$ 200 GeV is made. On
top of this cut, the experiment uses Boosted Decision Trees (BDT) cuts
based on rapidity differences between jets $\Delta \eta_{jj}$ and also
the gap between the $W$ boson rapidity and the average of the two
jets, cuts that are tailored to enhance VBF kinematics. We cannot
reproduce the BDT analysis based on public information, hence we
cannot do a reinterpretation of the data, but we can at least assess
whether those cuts would harm our signal to motivate a dedicated
search. We have checked that the signal has a very high acceptance to
the $m_{jj}$ and $\Delta \eta_{jj}$ cuts. On the other hand, the
signal has a low acceptance to the two other variables used in the
analysis, the so-called Zeppenfeld's variables
\cite{Schissler:2013nga}, rapidity difference $y^*=y_W-\frac{1}{2}
(y_{j_1}+y_{j_2})$ and the ratio $z^*=y^*/\Delta y_{jj}$, which are
not tailored to our situation, where one of the jets is coming with
the $W$ boson from the same heavy mother. Therefore, a dedicated
analysis based on the same variables as CMS's but with modified cuts
on the Zeppenfeld variables could have good sensitivity prospects for
this type of signal.

We mention in passing that for $Q_7$ the state with $Q=5/3$, once produced
at the LHC, would decay to $W+j$ and thus $Q_7$ could be constrained
in single VLQ searches in a similar way as discussed above for $Q_1$.

\subsection{Fake NTGCs: Cross sections and limits\label{subsect:PhenoB}}

\begin{figure}[t!]
    \centering
    \includegraphics[scale=0.72]{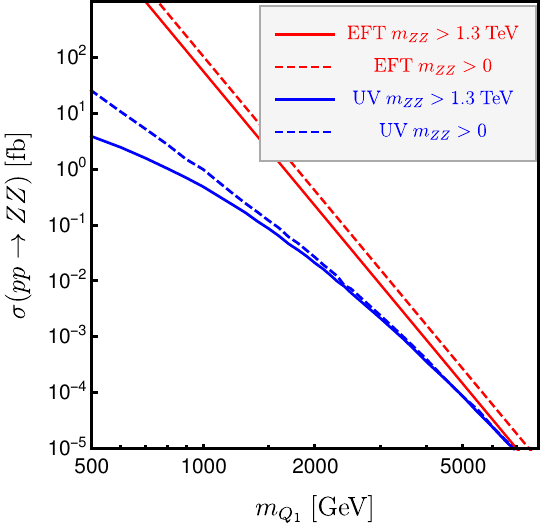}\hskip1mm
      \includegraphics[scale=0.45]{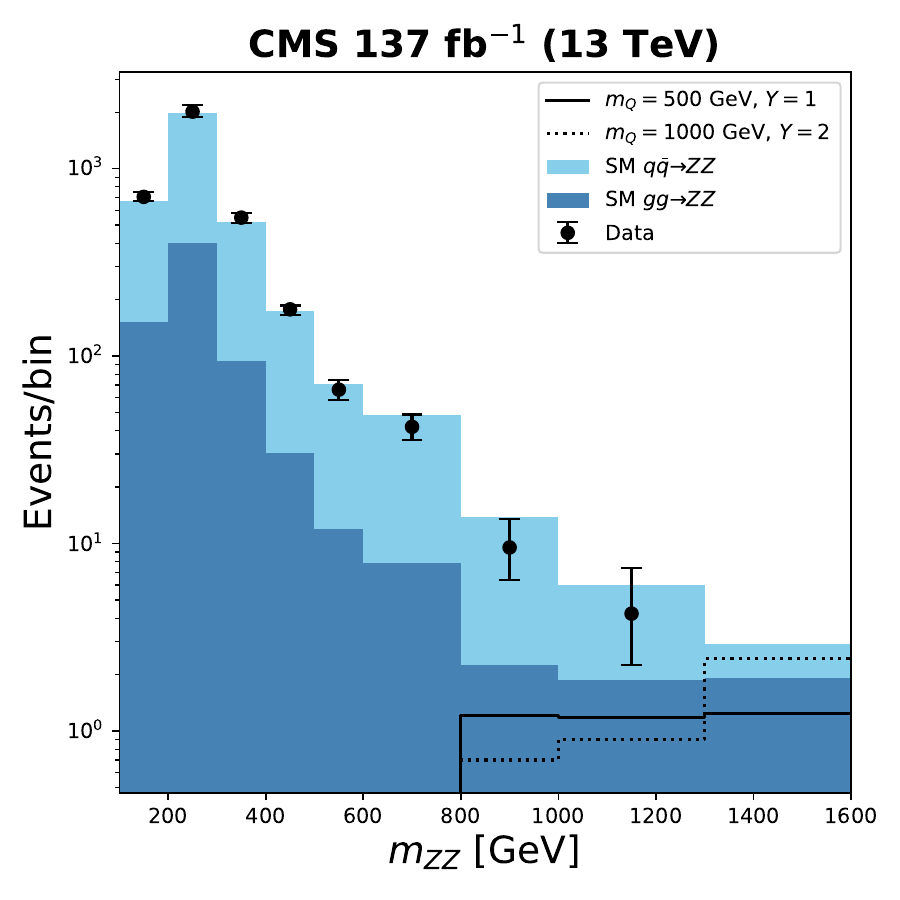}
    \caption{{\it Left panel: }LHC cross sections for $pp \to ZZ$ as
      function of $m_{Q_1}$ for $\sqrt{s}=14$ TeV. The red lines are
      the EFT calculation, the blue lines show the results of the
      cross section calculation from the UV model directly.  Results
      for two different cuts on $m_{ZZ}$ are shown. All calculations
      use $Y_{Q_u}^{(1)}=1$ (all other Yukawas equal to
      zero). Clearly, EFT calculation and UV model disagree below
      $m_{Q_1}\le 5$ TeV. {\it Right panel: }Invariant mass of the
      $ZZ$ system. The CMS \cite{CMS:2020gtj} SM background
      expectations are shown in blue, the measurements and their
      uncertainties are shown in black dots and bars, and the signal
      simulation in the last three bins is shown for two benchmarks, a
      light benchmark of $m_Q=$ 500 GeV and Yukawa equals 1, and a
      heavier mass $m_Q=1$ TeV with a larger Yukawa.}
    \label{fig:Xsect2}
\end{figure}

Turning now to pair production of $Z$-bosons,
Fig.\ (\ref{fig:Xsect2}), to the left, shows $\sigma(pp \to ZZ)$ as
function of $m_{Q_1}$ for different assumptions. The plot uses
$Y_{Q_u}^{(1)}=1$ (other Yukawas zero) and cross sections scale as the
4th power of the (sum of the) Yukawa couplings.  We have calculated
$\sigma(pp \to ZZ)$ twice, once with a direct implementation of the
VLQ model into \texttt{Feynrules} and once calculated from the
complete set of $d=8$ operators that the model generates at tree-level
in the limit of large $m_{Q_1}$. Each calculation is done twice, once
without a cut on $m_{ZZ}$ and once using $m_{ZZ}\ge 1.3$ TeV. Note
that the recent CMS search for NTGCs \cite{CMS:2020gtj} uses this cut
to eliminate standard model background.

As expected, the EFT and the UV model calculations agree only in the
limit of large values of $m_{Q_1}$. Nevertheless, we think it is worth
stressing that (approximate) agreement between the two calculations is
reached only for values of $m_{Q_1} \ge 5$ TeV or so. Such large
values of $m_{Q_1}$ are outside the sensitivity range of the
LHC. Thus, we must conclude that the EFT calculation does not provide
reliable results and seriously overestimates the cross section in the
LHC relevant mass range. For the extraction of limits on VLQ parameters
we will therefore use the UV model calculation.

The recent NTGC search by CMS \cite{CMS:2020gtj} can be used to derive
limits on VLQ parameters. In this search, two $Z$-bosons are tagged in
the final state and their invariant mass (a measure of the total
energy of the event) is used as a variable to capture new physics
effects. The CMS results are shown in the right panel of
Fig.\ \ref{fig:Xsect2}, where the experiment's SM predictions are
shown as blue distributions and the observed data as black dots with
bars corresponding to 1-$\sigma$ CL.
We also show our Monte Carlo predictions for two benchmark
signals in the last three bins. The lighter benchmark, $m_Q=$ 500 GeV
and $Y=1$, is rather flat with the increase of $m_{ZZ}$, whereas the
heavier benchmark of 1 TeV shows a relative growth with energy,
resembling more closely the EFT behaviour.

\begin{figure}[t!]
    \centering
      \includegraphics[scale=0.3]{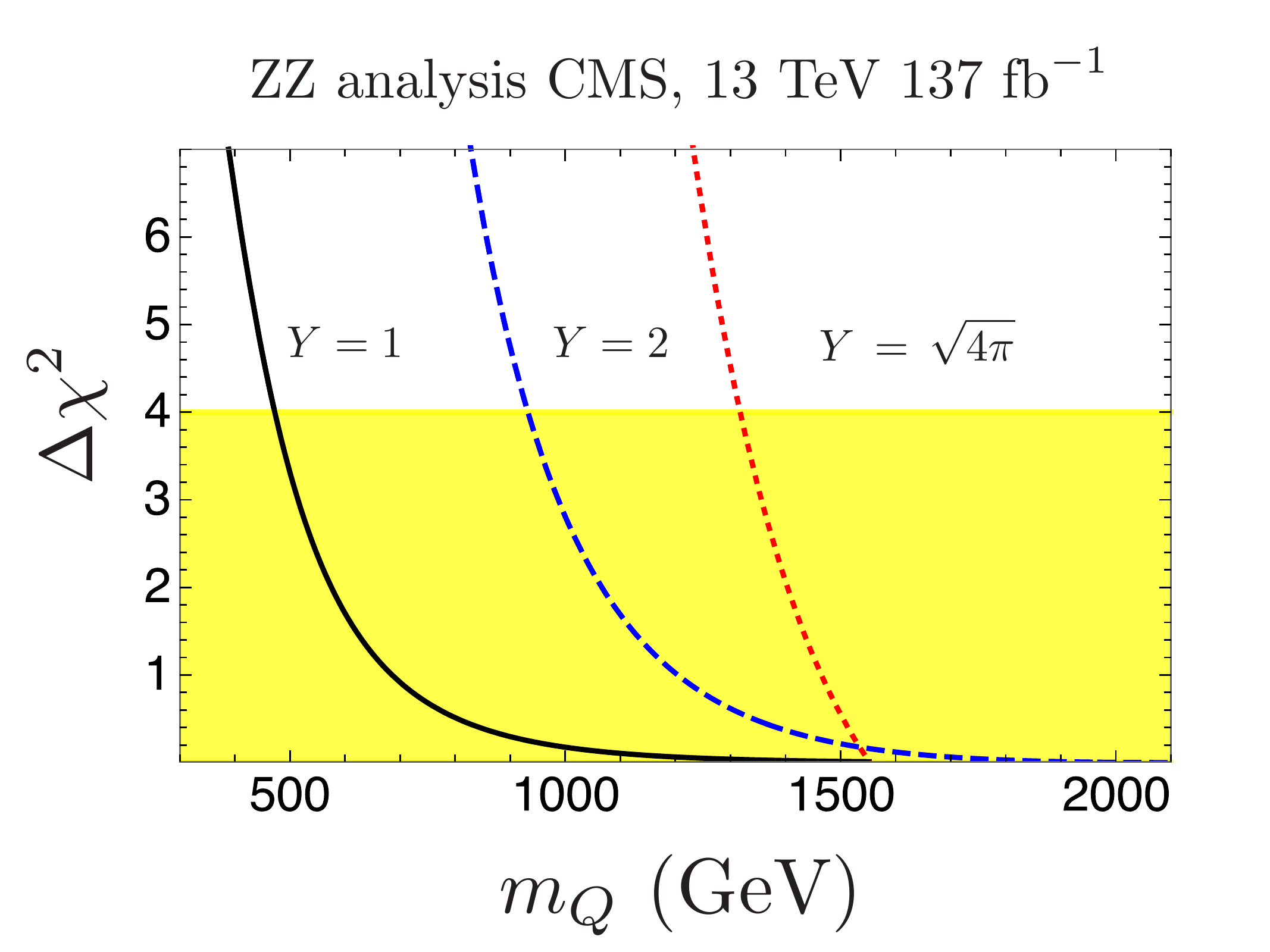}
     \caption{Binned analysis using the last three bins ($m_{ZZ}\in
       [800,1000,1300,\infty]$ GeV) from fig. (7) of
       \cite{CMS:2020gtj}.  Shown is the $\chi^2$ versus $m_{Q_1}$ for
       three different choices of $Y_{Q_u}^{(1)}$.}
    \label{fig:mQvsY}
\end{figure}

%MKH: I commented out this statement, since a more detailed discussion
%     is given below.
%After assessing that our MC simulation reproduces the $m_{ZZ}$ behaviour with o%rder one k-factors and mild dependence with energy, we can move onto using the %CMS data to set limits on the models. 

Fig. \ref{fig:mQvsY} shows the calculated $\chi^2$ as a function of
$m_{Q_1}$ for three different values of the Yukawa coupling
$Y_{Q_u}^{(1)}=1,2,\sqrt{4\pi}$. In this calculation, we used the last
3 bins in Fig.\ 7 of \cite{CMS:2020gtj} to extract the standard model
background. There are no events in the last bin,
i.e. $m_{ZZ}=[1.3,\infty]$ TeV, but the background simulation of CMS
estimates the SM background to be around $\sim 2.8$ events.

Given that the lower limit on $m_{Q_1}$ is around $m_{Q_1} \gsim 1.3$
TeV (see above) from pair production VLQ searches only a tiny region
of $m_{Q_1}$ with $Y_{Q_u}^{(1)}$ very close to the perturbative limit
can be constrained additionally from current NTGC searches. However,
as already stressed above, NTGC searches are not background limited,
while pair production VLQ searches are. Thus limits on VLQ parameter
space from NTGC searches should improve more rapidly with increasing
statistics than limits from pair produces VLQs.

\begin{figure}[t!]
    \centering
    \includegraphics[scale=0.7]{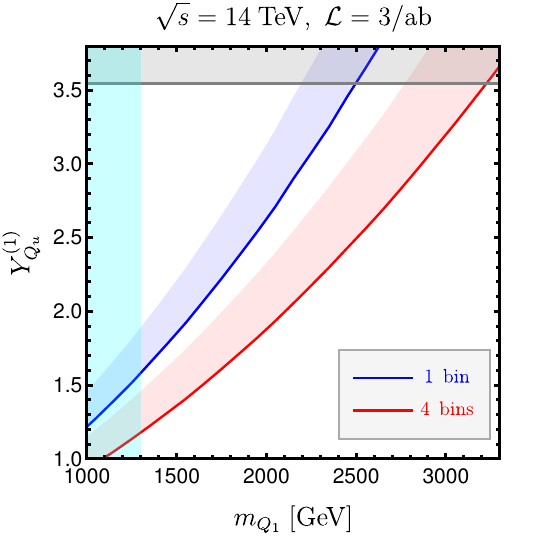}
    \includegraphics[scale=0.7]{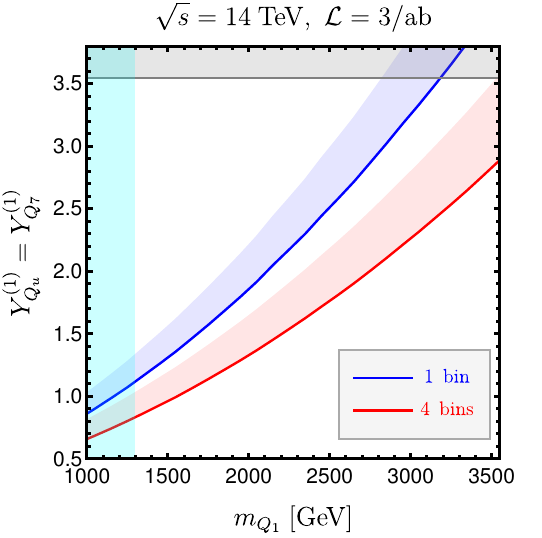}
    \caption{Sensitivity estimate for the high-luminosity LHC on VLQ
      parameter space using the on-shell $ZZ$ production search with
      leptonically decaying $Z$-bosons. The figure shows two cases. To
      the left: Only $Y_{Q_u}^{(1)}$ is non-zero; to the right:
      $Y_{Q_u}^{(1)}=Y_{Q_7}^{(1)}$. For discussion see text.}
    \label{fig:LimFut}
\end{figure}

In Fig.\ \ref{fig:LimFut} we present our estimate for the future
sensitivity of NTGC searches to constrain VLQ parameters. The
plots are calculated for $\sqrt{s}=14$ TeV and an assumed accumulated
luminosity of ${\cal L}=3/$ab. For the sensitivity estimate we use
the parameter $\alpha$, defined as \cite{Bityukov:2000tt}:
\begin{equation}
\alpha= 2 (\sqrt{S+B} - \sqrt{B}), \, 
\end{equation}
Here, $S$ is the number of signal events, while $B$ is the number of
background events. This formula reduces to the well-known
$S/\sqrt{B}$~\cite{Feldman:1997qc} in the limit $S\ll B$, but is more
robust under fluctuations and thus more suited to estimate
experimental prospects in regions with small numbers of
events~\cite{Barducci:2015ffa}.

Fig. (\ref{fig:LimFut}) shows our sensitivity estimate using $\alpha$
in the plane Yukawa coupling versus $m_{Q_1}$ for two different cases.
To the left: Only $Y_{Q_u}^{(1)}$ is non-zero. To the right:
$Y_{Q_u}^{(1)}=Y_{Q_7}^{(1)}$. The latter is motivated by the
``cancellation region'' for low-energy constraints discussed in
section \ref{sec:cnst}.  In order to calculate $\alpha$, however, we
need to estimate the number of background events for this hypothetical
future search. For this estimate we use a combination of cross section
calculation by \texttt{MadGraph} and the CMS data. We calculated
differential cross sections $\sigma(pp\to ZZ)$ at $\sqrt{s}=13$ TeV as
function of $m_{ZZ}$ in \texttt{MadGraph} and compare to the CMS
distribution, see fig. (7) in \cite{CMS:2020gtj} and also
fig. (\ref{fig:Xsect2}) to the right. We then normalize the
\texttt{MadGraph} distribution to the CMS one and use this ratio as a
correction factor as function of $m_{ZZ}$. With this fit function plus
the theoretical cross section we can ``predict'', as a cross-check for
this procedure, the number of expected background events in the bin
$m_{ZZ}=[2.0,\infty]$ TeV, which is not shown in the figure (and not
included in the derivation of the fit function), but mentioned in the
text. CMS estimates $\sim 0.2$ background events in this bin, our
estimate gives $\sim 0.27$. We then use the same function for
$\sigma(pp\to ZZ)$ at $\sqrt{s}=14$ TeV to estimate the expected SM
background events in each $m_{ZZ}$ bin for ${\cal L}=3$/ab.  Clearly,
this extrapolation procedure can be considered only as a rough
estimation of the future backgrounds and our sensitivity estimates
thus come with some error. For this reason in fig. (\ref{fig:LimFut})
we show our estimates as a line with an ``error'' band attached. This
error band represents the pessimistic scenario where no correction is
applied to the \texttt{MadGraph} output at all. Note that for this
pessimistic choice the limits on the Yukawa coupling, for a fixed
mass, are around 20\% weaker than for the case with correction
applied.

Note that each plot shows the result of 2 different calculations. The
line labelled ``1 bin'' calculates $\alpha$ in a single $m_{ZZ}$ bin
from $m_{ZZ}=[1.3,\infty]$. The line labelled ``4 bins'' divides the
calculation into more bins:
$m_{ZZ}=[1.3,1.6],[1.6,2.0],[2.0,2.5],[2.5,\infty]$.  Clearly, the
sensitivity improves considerably using more bins. This is easily
understood from the distributions shown in fig. (\ref{fig:SigmZZ}).
For the 4 bin calculation sensitivities in excess of $m_{Q_1}\gsim
3.3$ TeV are reached. However, all of the parameter space that can be
probed by this search is restricted to $Y_{Q_u}^{(1)} \gsim 1$.

Also note that the signal cross section $\sigma(pp\to ZZ)$ scales
proportional to the 4$^{\rm th}$ power of the Yukawa coupling, thus
for $m_{Q_1}$ at the current lower limit of $1.3$ TeV for
$Y_{Q_u}^{(1)}$ at the non-perturbative boundary around $\sim 450$
events are expected, while the sensitivity limit on $Y_{Q_u}^{(1)}$
at that mass of roughly $Y_{Q_u}^{(1)} \sim 1.2$ corresponds to an
upper limit of roughly 6 events.

In summary, the VLQ model can generate a non-zero number of
on-shell $ZZ$ events in the signal region used for NTGC searches
in a part of VLQ model parameter space that is currently not ruled
out by any low-energy or LHC search. If such events were indeed
seen in the future, single VLQ production searches should be
used by the experimental LHC collaborations to confirm or rule
out this alternative explanation for NTGC-like events.

\section{Conclusions\label{sec:cncl}}

In this work, we have calculated cross sections for on-shell pair
production of $Z$ bosons, $\sigma(pp\to ZZ)$, in a simple standard
model extension containing a vector-like quark doublet. Pairs of
on-shell $Z$ bosons have been used by both ATLAS \cite{ATLAS:2017bcd}
and CMS \cite{CMS:2020gtj} to derive limits on neutral triple gauge
bosons couplings (NTGCs). From the absence of NTGC like events in
these searches one can derive limits on the VLQ mass, $m_{Q_1}$,
as function of the VLQ Yukawa coupling.

Currently, these limits are rather weak and can constrain $m_{Q_1}$
above the pair production limit from \cite{ATLAS:2024zlo} only for
Yukawa couplings very close to non-perturbativity. However, direct VLQ
searches from pair production are background-limited, while in the
NTGC search of CMS \cite{CMS:2020gtj} there are zero events above
$m_{ZZ} = 1.3$ TeV. Thus, one can expect that the sensitivity to VLQ
parameters in such NTGC searches will increase (nearly) linearly with
increasing luminosity.

On the other hand, however, if NTGC searches based on on-shell $ZZ$
bosons will find an excess above SM background in the future, our work
shows that this observation can not unambiguously be interpreted as
the discovery of a neutral triple gauge boson vertex. For this reason
we have called the on-shell $ZZ$ boson production via a VLQ ``faking a
NTGC''. In such a -- admittedly optimistic -- case, further work will
be needed. We believe there are two possible experimental searches
that will allow to distinguish a ``real'' NTGC from a ``fake'' one.

The first possibility relies on the fact that the VLQ model requires
very large Yukawa couplings to generate a sizeable number of
on-shell $ZZ$ events. Single VLQ production thus has a large
cross section in that part of parameter space, where 
$\sigma(pp\to ZZ)$ is observable. Current single VLQ production
searches by ATLAS \cite{ATLAS:2022ozf} and CMS \cite{CMS:2024qdd}
do not provide limits for the relevant VLQ parameters, since
they both tag 3$^{\rm rd}$ generation quarks. An extension of these
searches to 1$^{\rm st}$ generation quarks should be able to test
the parameter space where VLQs can fake NTGCs. The current searches on single-bosons plus two jets, geared towards vector-boson fusion production, should be sensitive to our signal despite the low acceptance of some cuts used in the analysis, like the Zeppenfeld \cite{Schissler:2013nga} variables $y^*$ and $z^*$. Unfortunately, the use of BDTs for selection and the lack of public information on how to implement them, prevents us from producing an estimate of the current sensitivity. Nevertheless, a dedicated analysis with a single $W$, $Z$ or Higgs boson with two very energetic jets, and with a mass bump in a combination of  boson and jet should be distinctive enough as to motivate a dedicated search.   

The second possibility to distinguish fake and ``real'' NTGCs is to
use the final state $Z\gamma$. There is a CMS search for this final
state based on $\sqrt{s}=13$ TeV and ${\cal L}=138/$fb
\cite{CMS:2024tag}. The limits on $h_3^{\gamma / Z}$ from this study
show that competitive limits, when compared to the limits on
$f_5^{\gamma /Z}$ from \cite{CMS:2024qdd}, can be obtained in such
a search. Observing $Z\gamma$ would indeed be a ``proof'' for
the existence of NTGCs, since the VLQ model can not fake
this final state in the same way as for $ZZ$.

\section*{Acknowledgements}

We would like to thank Javier Fuentes-Martin, Julie Pages and Anders
Eller Thomsen for help with \texttt{Matchete}
\cite{Fuentes-Martin:2022jrf}. F.E. thanks Peter Stangl for help with
\texttt{flavio} \cite{Straub:2018kue} and Eleftheria Solomonidi for
discussions about meson decays.  F.E. is supported by the Generalitat
Valenciana under the grants GRISOLIAP/2020/145 and
PROMETEO/2021/083. M.H. acknowledges support by grants
PID2020-113775GB-I00 (AEI/10.13039/ 501100011033) and CIPROM/2021/054
(Generalitat Valenciana).  R.C. is supported by
MCIN/AEI/10.13039/501100011033 and the European Union
NextGenerationEU/PRTR under the grant JDC2022-048687-I and partially
funded by Grant AST22\_6.5 (Consejeria de Universidad, Investigacion e
Innovacion and Gobierno de Espa\~na and Union Europea
NextGenerationEU). V.S. is supported by the Generalitat Valenciana
PROMETEO/2021/083, Proyecto Consolidacion CNS2022-135688, and the
Ministerio de Ciencia e Innovacion PID2020 -113644GB-I00.

\bigskip

\appendix
\section{Appendix: 1-loop NTGCs in the $Q_1$ model\label{sect:app}}
\label{sec:appNTGC}

As shown in \cite{Cepedello:2024ogz} there are four $d=8$ SMEFT
operators, that contribute to NTGCs with two on-shell boson. They are
defined as:
\begin{eqnarray} 
\label{eq:ODBB}
{\cal O}_{DB\tilde{B}} &=& i \frac{c_{DB\tilde{B}}}{\Lambda^4}
     H^{\dagger} {\tilde B_{\mu\nu}} (D^{\rho} B_{\nu\rho}) D_\mu H + {\rm h.c.} \, , 
      \\
\label{eq:ODWW}
{\cal O}_{DW\!\tilde{W}} &=& i  \frac{c_{DW\tilde{W}}}{\Lambda^4}
     H^{\dagger} {\tilde W_{\mu\nu}} (D^{\rho} W_{\nu\rho}) D_\mu H + {\rm h.c.} \, , 
     \\
\label{eq:ODWB}
{\cal O}_{DW\!\tilde{B}} &=& i  \frac{c_{DW\tilde{B}}}{\Lambda^4}
     H^{\dagger} {\tilde B_{\mu\nu}} (D^{\rho} W_{\nu\rho}) D_\mu H + {\rm h.c.} \, ,
     \\
\label{eq:ODBW}
{\cal O}_{DB\tilde{W}} &=& i  \frac{c_{DB\tilde{W}}}{\Lambda^4}
     H^{\dagger} {\tilde W_{\mu\nu}} (D^{\rho} B_{\nu\rho}) D_\mu H + {\rm h.c.} \, .
\end{eqnarray}
These four operators contribute to the form factors $f_5^{Z/\gamma}$
and $h_3^{Z/\gamma}$, that parametrize the vertices $ZZZ^*$/$ZZ\gamma^*$
and $Z\gamma Z^*$/$Z\gamma\gamma^*$, as defined in \cite{Gounaris:2000tb},
via:\footnote{NTGCs vanish if all three gauge bosons are on-shell, the ``*''
indicates the off-shell boson}
\begin{eqnarray}
\label{eq:f5Z}
f_5^Z & = & \frac{v^2m_Z^2}{\Lambda^4} \, \frac{1}{c_W s_W} 
  \left[ s_W^2 c_{DB\tilde{B}} + c_W^2 c_{DW\!\tilde{W}} + \frac{1}{2} c_W s_W (c_{DW\!\tilde{B}} + c_{DB\tilde{W}}) \right],
  \\
\label{eq:f5Gam}
f_5^\gamma & = & \frac{v^2m_Z^2}{\Lambda^4} \, \frac{1}{c_W s_W} 
  \left[ c_W s_W ( - c_{DB\tilde{B}} + c_{DW\!\tilde{W}} ) - \frac{1}{2} (s_W^2 c_{DW\!\tilde{B}} - c_W^2 c_{DB\tilde{W}} )  \right],
    \\
\label{eq:h3Z}
  h_3^Z & = & \frac{v^2m_Z^2}{\Lambda^4} \, \frac{1}{c_W s_W} 
  \left[ c_W s_W ( - c_{DB\tilde{B}} + c_{DW\!\tilde{W}} ) + \frac{1}{2} (c_W^2 c_{DW\!\tilde{B}} - s_W^2 c_{DB\tilde{W}} )  \right],
  \\
\label{eq:h3Gam}
  h_3^\gamma & = & \frac{v^2m_Z^2}{\Lambda^4} \, \frac{1}{c_W s_W} 
  \left[ c_W^2 c_{DB\tilde{B}} + s_W^2 c_{DW\!\tilde{W}} - \frac{1}{2} c_W s_W ( c_{DW\!\tilde{B}} + c_{DB\tilde{W}} ) \right] ,
\end{eqnarray}
where $c_W$ and $s_W$ are the cosine and sine of the weak-mixing
angle.

The standard model extension with a $Q_1$ VLQ contributes to all
four operators, eqs (\ref{eq:ODBB})-(\ref{eq:ODBW}), at the level
of 1-loop. We have used  \texttt{Matchete} \cite{Fuentes-Martin:2022jrf}
to calculate the matching and find:
\begin{eqnarray}
\label{eq:cdbtb}
c_{DB\tilde{B}} = \frac{-7i}{72 (16 \pi^2)}g_Y^2 \Big(|Y_{Q_d}|^2 -|Y_{Q_u}|^2 \Big)
\\ 
\label{eq:cdwtw}
c_{DW\tilde{W}} = \frac{-i}{24 (16 \pi^2)}g_2^2 \Big(|Y_{Q_d}|^2 -|Y_{Q_u}|^2 \Big)
\\ 
\label{eq:cdwtb}
c_{DW\tilde{B}} = \frac{i}{12 (16 \pi^2)}g_2g_Y \Big(|Y_{Q_d}|^2 -3|Y_{Q_u}|^2 \Big)
\\ 
\label{eq:cdbtw}
c_{DB\tilde{W}} = \frac{-i}{12 (16 \pi^2)}g_2g_Y \Big(5|Y_{Q_d}|^2 -9|Y_{Q_u}|^2 \Big)
\end{eqnarray}
Inserting numerical values for couplings constants and for
$Y_{Q_d}\equiv 0$ and $Y_{Q_u}^{(1)}=1$ and $\Lambda =1$ TeV, this
leads to $f_5^{Z} \simeq 2.1 \times 10^{-7}$, $f_5^{\gamma} \simeq 2.7
\times 10^{-7}$, $h_3^{Z} \simeq -8.5 \times 10^{-8}$ and
$h_3^{\gamma} \simeq -5.15 \times 10^{-8}$. These numbers should be
compared to the current upper limits
$f_5^{Z/\gamma} \lsim 7.5 \times 10^{-4}$ \cite{CMS:2020gtj}  and
$h_3^{Z} \simeq 2.2 \times 10^{-4}$  and 
$h_3^{\gamma} \simeq 3.5 \times 10^{-4}$ \cite{CMS:2024tag}.

\bibliographystyle{JHEP}
\bibliography{FakeZZZ}

\end{document}